\documentclass[10pt,leqno]{amsart}
\usepackage{graphicx}
\usepackage{textcase}
\usepackage{indentfirst,csquotes}

\usepackage{amssymb,amsthm,amsmath}
\usepackage{bm}
\usepackage{xcolor,paralist,hyperref,fancyhdr,etoolbox}
\usepackage{orcidlink}
\usepackage{framed}
\usepackage{fancyvrb}
\usepackage[export]{adjustbox}
\setkeys{Gin}{max width=\textwidth}

\hypersetup{colorlinks=true,linkcolor=black,citecolor=black,filecolor=black,urlcolor=black}

\newenvironment{CodeChunk}{}{}
\DefineVerbatimEnvironment{CodeInput}{Verbatim}{fontsize=\small}

\makeatletter

\renewcommand{\section}{%
  \@startsection{section}{1}%
  {\z@}%
  {1.5\baselineskip}%
  {0.8\baselineskip}%
  {\normalfont\Large\bfseries\centering}%
}

\renewcommand{\subsection}{%
  \@startsection{subsection}{2}%
  {\z@}%
  {1.2\baselineskip}%
  {0.5\baselineskip}%
  {\normalfont\large\bfseries\centering}%
}

\makeatother

\begin{document}
\title{\NoCaseChange{Visualizing Class Specific Heterogeneous Tendencies using \textsf{R}}}

\author[Takagishi and van de Velden]{
\NoCaseChange{Mariko Takagishi}~\orcidlink{0000-0002-2984-8991}\\
{\small\normalfont\NoCaseChange{Faculty of Engineering, Mathematical and Data Sciences Program}}\\
{\small\normalfont\NoCaseChange{Graduate School of Environmental and Life Science}}\\
{\small\normalfont\NoCaseChange{Okayama University}}\\
\mbox{}\\
\NoCaseChange{Michel van de Velden}~\orcidlink{0000-0002-9807-9057}\\
{\small\normalfont\NoCaseChange{Econometric Institute}}\\
{\small\normalfont\NoCaseChange{Erasmus School of Economics}}\\
{\small\normalfont\NoCaseChange{Erasmus University Rotterdam}}
}

\thanks{Corresponding author: Mariko Takagishi (\texttt{takagishi@okayama-u.ac.jp})}

\date{\today}
\maketitle

\begin{abstract}
In this paper we introduce the \textsf{R} package \textbf{mccca}, which implements multiple-class cluster correspondence analysis (MCCCA) proposed in \cite{1}. MCCCA is a statistical method that identifies and visualizes heterogeneous tendencies specific to ``classes'' (e.g., gender and nationality) in a low dimensional space. In MCCCA, two kinds of variables, external and active variables, are distinguished. External variables directly define classes whereas the active variables are used to derive class-specific clusters of observations (individuals). In this paper, we show how to apply MCCCA and how to visualize its results using \textbf{mccca}. We illustrate the procedure for performing MCCCA by applying it to two data sets.
\end{abstract}

\bigskip
%
\section{Introduction} \label{sec:intro}


Correspondence analysis (CA) and multiple correspondence analysis (MCA) are popular methods for the analysis of categorical data (e.g., \cite{2}). In these methods, rows and columns of a data matrix (for CA; a contingency matrix, for MCA a super-indicator matrix) are plotted in a low-dimensional space, allowing visual interpretation of the relationships among and between rows and columns. In particular, CA or MCA results can be visualized as a biplot in which distances and projections can be interpreted. See, e.g., \cite{3}. 

Adding external information about observations can enhance the interpretation of MCA biplots. One approach for incorporating external information into MCA was proposed by \cite{4}. An alternative approach, multiple-class cluster correspondence analysis (MCCC), allowing for class specific differences, was proposed in \cite{1}. Here we consider an \textsf{R} package implementing MCCCA and corresponding visualization methods. MCCCA is a method that incorporates external information that divides observations into known ``classes'', e.g., gender, nationality) and aims to simultaneously visualize clusters of observations and variable associations, contingent on these known classes. 

A straightforward way to incorporate external information, that we shall refer to as the averaging approach, is to consider class-specific averages. However, by doing so, heterogeneous tendencies within a class are ignored. For example, suppose that within a class (e.g., males) a majority exhibits a strong preference (tendency) towards a particular category of a certain variable, while preferences in other, smaller groups, differ consistently. In this case, in the averaging approach, the strong tendency of the larger group will dominate the results and obscure the preferences in the smaller groups. MCCCA overcomes this problem by distinguishing between two kinds of variables: external and active variables. The external variables are used to define classes, whereas the active variables are used to obtain class-specific clusters of observations exhibiting similar tendencies.

In this paper, we describe an R package, \textbf{mccca}, that implements the main procedures described in \cite{1}. Moreover, we show how the package can be used for model selection, and how results can be summarized and visualized. We illustrate the procedures for applying and interpreting MCCCA using two empirical data sets available in the package.

The outline of the paper is as follows. In Section 2, we briefly review MCCCA as proposed in \cite{1} and introduce some notation. Next, in Section 3, we illustrate an overview of the \textsf{R} package \textbf{mccca} available from the Comprehensive R Archive Network (CRAN) at \url{https://CRAN.R-project.org/package=mccca}. Two example data sets are used to illustrate various additional estimation and visualization options. In Section 4, limitations and possible extensions of \textbf{mccca} are discussed.

%
\section{MCCCA: Multiple-Class Cluster Correspondence Analysis} \label{sec:models}
%

Suppose that we have data consisting of $N$ observations on $J$ categorical variables. Furthermore, for the same $N$ observations, we have $H$ additional categorical variables that can be used to distinguish subgroups. Throughout this paper, we refer to the two sets of categorical variables as ``active'' (the first $J$) and ``external'' (the additional $H$) variables, respectively. The $H$ external variables are used to define $C$ classes by considering either specific categories of the external variables, or combinations (i.e., interactions) of some or all of these categories. We can schematically represent this type of data using a two-level hierarchical structure in which the external categories are used to split the data into $C$ \emph{known} ``classes'', and within each class, observations belong to several \emph{unknown} clusters, called ``class-specific clusters''. Application of MCCCA results in a simultaneous visualization of the class-specific clusters (representing the heterogeneous tendencies) and the categories of the active variables. This joint representation allows for an interpretation of the class-specific tendencies as well as the relationship between the clusters and the categories of the active variables. 

We can express MCCCA as a constrained minimization problem. To do so, we introduce the following notation:
\begin{table}[hbt!]
\centering
\caption{{\small Some notations in MCCCA}}
\scalebox{0.85}{
\begin{tabular}{@{}l p{0.64\textwidth}@{}}
\hline
Notation & Description\\
\hline
$N$ & number of observations\\
$J$ & number of active (categorical) variables\\
$H$ & number of external variables\\
$q_j$, $(j=1,\ldots,J)$ & number of categories for the $j$th active variable\\
$Q=\sum_{j=1}^Jq_j$ & total number of categories among $J$ active categorical variables\\
$C_h$, $(h=1,\ldots,H)$ & number of categories for the $h$th external variable\\
$C=\prod_{h=1}^HC_h$ & total number of classes (e.g., if $C_1=2, C_2=2$, $C=4$)\\
$c_h$, $(c_h=1,\ldots,C_h;$ & index for the categories of the $h$th external variable\\
$\qquad\quad h=1,\ldots,H)$ & (i.e., $c_1=1$ and $c_2=1$ indicate the classes corresponding to the category ``1'' of the 1st and 2nd external variables respectively)\\
$K_{c_1\cdots c_H}$ & number of clusters for the $(c_1,\ldots,c_H)$ class\\
$K=\sum_{c_1=1}^{C_1}\cdots\sum_{c_H=1}^{C_H}K_{c_1\cdots c_H}$ & total number of clusters among all classes\\
$p$ & number of dimensions\\
$\bm{Z}_j$, $(j=1,\ldots,J)$ & $N\times q_j$ category indicator matrix for the $j$th active variable\\
$\bm{V}_h$, $(h=1,\ldots,H)$ & $N\times C_h$ category indicator matrix for the $h$th external variable\\
$\bm{B}_j$, $(j=1,\ldots,J)$ &
$q_j\times p$ quantification matrix for the categories of the $j$th active variable\\
$\bm{U}=(\bm{U}_{11\cdots1},\ldots,\bm{U}_{C_1C_2\cdots C_H})$
&
$N\times K$ cluster indicator matrix, where each
$\bm{U}_{c_1c_2\cdots c_H}$ is an
$(N\times K_{c_1c_2\cdots c_H})$ matrix\\
$\bm{G}=
\left(
\begin{array}{c}
\bm{G}_{11\cdots1}\\
\vdots\\
\bm{G}_{C_1C_2\cdots C_H}
\end{array}
\right)$
& $K\times p$ quantification matrix for cluster centers, where each
$\bm{G}_{c_1c_2\cdots c_H}$ is a $(K_{c_1c_2\cdots c_H}\times p)$ matrix\\
\hline
\end{tabular}
}
\label{tab:notation}
\end{table}

The objective function of MCCCA is defined as;
\begin{gather}
 	\min_{\bm{U},\bm{G},\bm{B}_j}\phi(\bm{U}, \bm{G}, \bm{B}_j\,|\,\bm{Z}_{j},\bm{V}_h)=\frac{1}{NJ}\sum_{j=1}^J\|\bm{U}\bm{G}-\bm{Z}_{j}\bm{B}_j\|^2\label{eq:obj}\\
 	{\rm s.t.}\quad\frac{1}{NJ}\sum_{j=1}^J\bm{B}_j^{\prime}\bm{Z}_{j}^{\prime}\bm{Z}_{j}\bm{B}_j=\bm{I}_p,\quad\bm{M}_{N}\bm{U}\bm{G}=\bm{U}\bm{G}\nonumber\\
 	{\rm where}\,\,\,\underset{(N\times K)}{\bm{U}}=\left(\bm{U}_{11\cdots 1},\bm{U}_{11\cdots 12},\ldots,\bm{U}_{11\cdots 1C_H},\ldots,\bm{U}_{C_1C_2\cdots C_{H-1},C_H}\right)\label{eq:U}%
\end{gather}

Here, $\bm{M}_{N}=\bm{I}_{N}-N^{-1}\bm{1}_{N}\bm{1}_{N}^{\prime}$ is the centering matrix, $\bm{I}_{N}$ is an $N\times N$ identity matrix, and $\bm{1}_{N}$ is an $N\times 1$ vector of ones. 

The unknown parameters in MCCCA are $\bm{U}$: the class-specific clusters defined by the cluster indicator matrix, $\bm{B}_j$, $(j=1,\ldots,J)$: the coordinates for the categories of the $J$ active variables, and $\bm{G}$: the coordinates for the class-specific cluster centers. To estimate these parameters, the number of class-specific clusters $K_{c_1\cdots c_H}$, and the dimensionality of the solution $p$, must be fixed. Then, conditional on these choices, parameters can be estimated using an alternating least squares algorithm. For more details on the algorithm and its properties see \cite{1}.

\begin{figure}[t!]
\centering
\includegraphics[width=12cm]{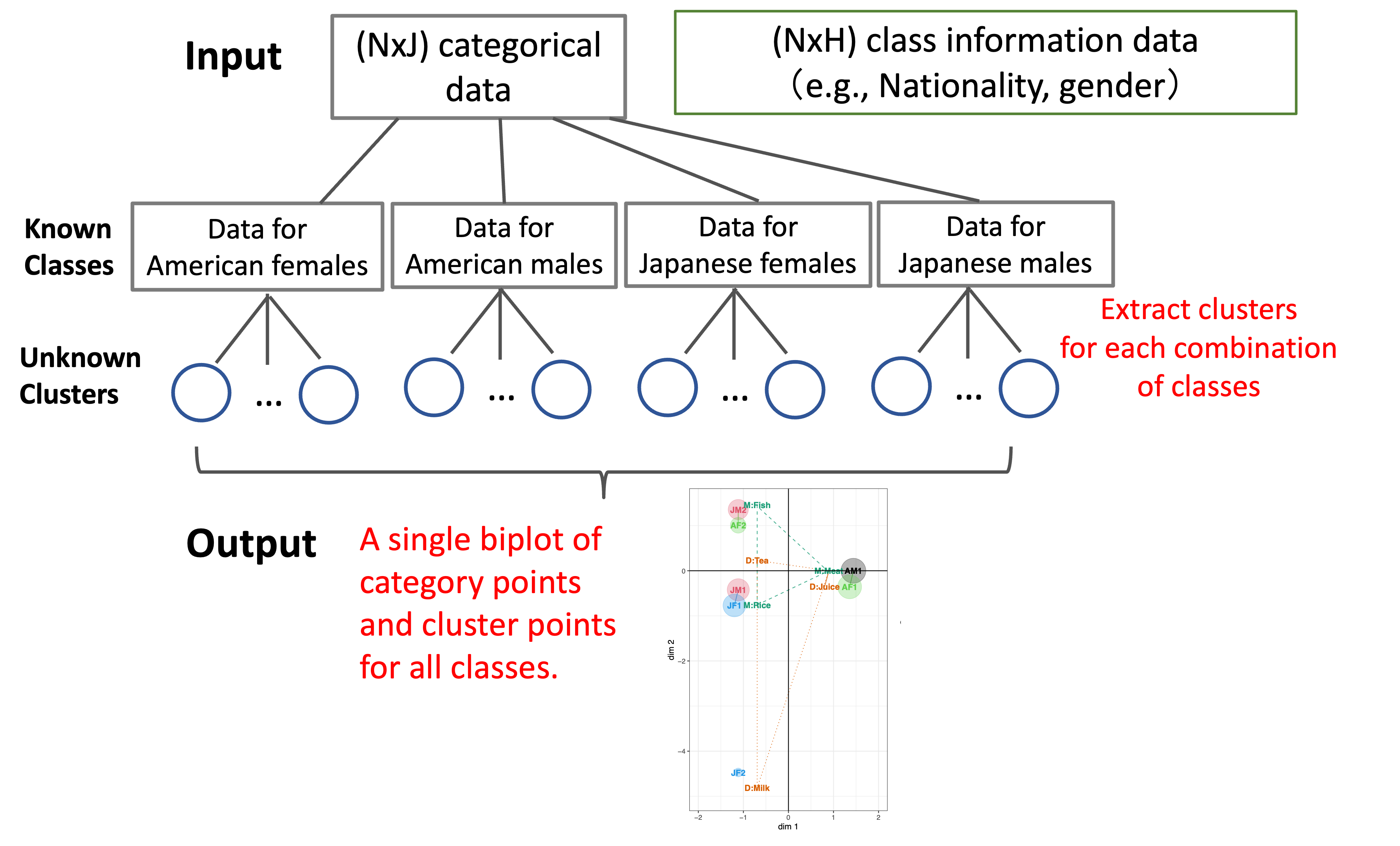}
\caption{\label{fig:MCCCA} Schematic overview of MCCCA for the meal preference example}
\end{figure}

Note that a two-level hierarchical structure is defined on $\bm{U}$, with known classes in the first level and unknown clusters for each class in the second level. In particular, it is important to observe that each combination of the categories of the external variables, defines a class for which clusters need to be obtained. For example, suppose that we have two external variables: nationality (e.g., American and Japanese) and gender (e.g., male and female), respectively. In this case, the number of categories for each external variable is two. Hence, $C_1=C_2=2$. Consequently, there are $4$ known classes for which clusters need to be obtained. Let $c_1=1$ and $c_1=2$ correspond, respectively, to the ``American'' and ``Japanese'' classes for the nationality variable, while $c_2=1$ indicates the ``female'' and $c_2=2$ the ``male'' class for the gender variable. Then, the ``$c_1=2, c_2=1$'' class indicates the combined ``Japanese female'' class, and the corresponding cluster indicator matrix $\bm{U}_{21}$ defines the clusters for this class. See Fig. $\ref{fig:MCCCA}$ for a visualization of the data and the resulting hierarchy for this small example. 

The number of clusters for each known class is defined by $K_{c_1\cdots c_H}$. Hence, for this example, $K_{21}=3$ indicates that there are $3$ clusters in the ``Japanese female class''. Similarly, $K_{11}$, $K_{12}$ and $K_{22}$ indicate the number of clusters for the ``American female'', ``American male'', and the ``Japanese male'' classes, respectively. Note that the numbers of (unknown) class-specific clusters have to be specified prior to analysis. For more technical details, see \cite{1}.

%
\section{Package description and illustrative examples
} \label{sec:illustrations}
%

In this section, we describe the main functions included in the package and we show, using two example data sets available in the package, how to perform MCCCA using the \textbf{mccca} package. 

In order to apply MCCCA, categorical data are required. Moreover, the data should consist of two types of variables: active and external variables. The external variables, and interactions between them, define the known classes. The total number of, and the distribution over, these classes may not always be obvious. Hence, the general flow of function usage is designed to first transform the data and inspect these dimensions. In particular, first, the data set needs to be transformed to an \texttt{mcccadata} object using \texttt{create.MCCCAdata()}. Next, \texttt{MCCCA()} is applied to the \texttt{mcccadata} object after which the results of \texttt{MCCCA()} can be visualized using \texttt{plot()}. A numerical summary can be obtained using \texttt{summary()}. To specify an appropriate number of clusters for each class, a dedicated tuning function \texttt{decideK()} can be used. Table $\ref{tab:functions}$ shows a summary of the package contents. We illustrate how the \textbf{mccca} package can be used by applying it to two datasets available in the package. In particular, we first consider an artificial data set, with known underlying cluster structure, to show the basic functionalities. Next, we use empirical data concerning road accidents in the UK, to illustrate more advanced functionalities. 

\begin{table}[t]
	\centering
	\caption{{\small Summary of functions in \textbf{mccca} package.}}
	\scalebox{0.8}{ 
			\begin{tabular}{ll}
				\hline
				    Function & Description\\
                    \hline
				\texttt{create.MCCCAdata()} & Creates \texttt{mcccadata} object which is later applied to \texttt{MCCCA()}.\\
				\texttt{MCCCA()} & Applies MCCCA to the \texttt{mcccadata} object created by \texttt{create.MCCCAdata()}.\\
                    \texttt{decideK()} & Calculates the cluster index (KL index by default) \\
                    & to determine the number of class-specific clusters.\\
                    \texttt{summary()} & Produces a detailed output for \texttt{create.MCCCAdata()}, \texttt{MCCCA()} and \texttt{decideK()}\\
				\texttt{plot()} & Shows a biplot using quantifications estimated by \texttt{MCCCA()}.\\
				\hline
			\end{tabular}
	}
	\label{tab:functions}
\end{table}

\subsection{Example: Meal and drink preferences}
\label{mealDrink}

The \texttt{mealDrink} data, described in Section 3 of  \cite{1} and available in the package, consists of artificial data for $390$ observations on $4$ categorical variables. Table $\ref{tab:mealDrink}$ shows the variables and corresponding categories in \texttt{mealDrink}. For these data, the aim is to identify whether there are different tendencies in meal and drink preferences (the active variables) depending on nationality and gender (the external variables). Before loading and inspecting the data, we install and load the  \texttt{mccca} package:
\begin{CodeChunk}
\begin{CodeInput}
R> install.packages("mccca")
R> library(mccca)
R> data(mealDrink)
\end{CodeInput}
\end{CodeChunk}

As we have two binary external variables ($C_1=C_2=2$), four classes ($C=C_1C_2=4$) can be made. These classes correspond to American females, American males, Japanese females, and Japanese males. The characteristics of these data are as follows: most Americans prefer Meat and Juice, while a smaller group of American females select Fish and Tea. Similarly, the majority of Japanese prefer Rice and Tea, while a small group of the Japanese females chooses Rice and Milk. Instead of only focusing on majority preferences, MCCCA allows us to visualize heterogeneous tendencies for the subgroups defined by the categories of the external variables. 

\begin{table}[t]
	\centering
	\caption{{\small Categories of variables of \texttt{mealDrink} data}}
	\scalebox{0.9}{ 
		\begin{tabular}{lll}
			\hline
			Variable type & Variable name & Category\\
			\hline
			Active variables & Meal & Meat, Fish, Rice \\
			& Drink & Fruits juice, Tea, Milk\\
			External variables & Nationality & American, Japanese \\
			& Gender & Female, Male\\
			\hline
		\end{tabular}
	}
	\label{tab:mealDrink}
\end{table}

Note that the mealDrink data set in \textbf{mccca}, consists of four categorical variables. The first two variables are active and the last two are external. To apply \texttt{MCCCA}, separate dataframes consisting of the active and external variables are required. We can create these as follows:
\begin{CodeChunk}
\begin{CodeInput}
R> active<-c(1,2)
R> external<-c(3,4)
R> dat.act=mealDrink[,active]
R> dat.ext=mealDrink[,external]
\end{CodeInput}
\end{CodeChunk}
Here, \texttt{dat.act} is an $(N\times J)$ dataframe with $J=2$ active variables, while \texttt{dat.ext} is an ($N\times H$) dataframe with $H=2$ external variables ($N=390$). The first rows of these data are:

\begin{CodeChunk}
\begin{CodeInput}
R> head(dat.act)
     meal  drink
obs1 Meat Fjuice
obs2 Fish    Tea
obs3 Meat Fjuice
obs4 Rice    Tea
obs5 Rice    Tea
obs6 Rice    Tea

R> head(dat.ext)
     nationality gender
obs1    American   male
obs2    Japanese   male
obs3    American female
obs4    Japanese female
obs5    Japanese female
obs6    Japanese   male
\end{CodeInput}
\end{CodeChunk}
Note that, when merging active and external variables from separate sources, the rows of the two dataframes should correspond. 

To apply MCCCA, we need to construct the classes as defined by the external variables. Rather than having this done internally in the \texttt{MCCCA} function, this needs to be done by applying the \texttt{create.MCCCAdata()} function. In that way, before applying MCCCA, the user is presented with an overview of the constructed classes as well as a distribution over these classes. 

\begin{CodeChunk}
\begin{CodeInput}
R> mcccadata=create.MCCCAdata(dat.act,dat.ext)
1th class: (American,female) data, n=105
2th class: (American,male) data, n=90
3th class: (Japanese,female) data, n=80
4th class: (Japanese,male) data, n=115
The total number of classes: 4 
\end{CodeInput}
\end{CodeChunk}
The data corresponding to the active and external variables are required arguments in \\
\texttt{create.MCCCAdata()}. By applying \texttt{create.MCCCAdata()}, all classes (corresponding to the combinations of categories of the external variables) are created, and the numbers of observations belonging to each class are printed. Furthermore, the function results in an object, here \texttt{mcccadata}, which is a list consisting of the data set split by class, and the information for each class. Table $\ref{tab:createMCCCAfunc}$ provides a complete list of all \texttt{create.MCCCAdata()} arguments and output. An overview of the resulting object can be obtained using the \texttt{summary()} function:

\begin{CodeChunk}
\begin{CodeInput}
R> summary(mcccadata)
The number of classes created by the given external variables: 4 

Class labels and its constituent external variables:
          nationality gender   classlabel       
1th class "American"  "female" "American&female"
2th class "American"  "male"   "American&male"  
3th class "Japanese"  "female" "Japanese&female"
4th class "Japanese"  "male"   "Japanese&male"  

The number of observations in each class:
American&female   American&male Japanese&female   Japanese&male 
            105              90              80             115 
\end{CodeInput}
\end{CodeChunk}

\begin{table}[t]
	\centering
	\caption{{\small List of \texttt{create.MCCCAdata()} arguments and outputs.}}
	\scalebox{0.75}{ 
			\begin{tabular}{ll}
				\hline
				    Arguments & Description\\
                    \hline
				\texttt{dat.act} & An ($N\times J$) data frame (matrix) of active categorical variables.\\
    & If \texttt{rownames(dat.act)} is \texttt{NULL}, \texttt{c(obj1,..,objN)} are used as \texttt{rownames(dat)}.\\
				\texttt{dat.ext} & An ($N\times H$) data frame (matrix) of external variables.\\
				\hline
    Output & Description\\
    \hline
    \texttt{C} & The number of classes created from the external variables.\\
    \texttt{dat.act} & A copy of \texttt{dat.act} in the input.\\
    \texttt{act.list} & A list of $C$ data frames, each of which is obtained by dividing \texttt{dat.act} into $C$ classes.\\
    \texttt{N.vec} & An integer vector of length $C$ giving the number of observations\\
    & belonging to each of the $C$ classes.\\
    \texttt{q.vec} & An integer vector of length $J$ giving the number of categories in each of the $J$ active variables.\\
    \texttt{classlabel}& A character vector of length $C$ giving the class labels for each of the $C$ classes.\\
    \texttt{class.label.n.vec} & A character vector of length $N$ giving the class labels for each observation.  The order is\\
    & the same as the order of the rows of \texttt{dat.act}\\
    & (i.e., \texttt{names(class.label.n.vec)}=\texttt{rownames(dat.act)}).\\
    \texttt{class.index.n.vec}& An integer vector of length $N$, giving the same information as \texttt{class.label.n.vec} \\ 
    & but represented by class index.\\ 
    \texttt{obs.index.n.list}& A list of $C$ integer vectors, with the observation indices corresponding to each of the $C$ classes.\\
    \texttt{classlab.mat}& A ($C \times (H+1)$) table showing the correspondence between each class and the external variables. \\
    & The rows indicate the classes, the first $H$ columns indicate the external category of\\
    & each external variable that constitutes a class, and the last column indicates the class label.\\
                    \hline
			\end{tabular}
	}
	\label{tab:createMCCCAfunc}
\end{table}

To apply MCCCA, the number of class-specific clusters needs to be specified using an integer vector of the appropriate length. This total number of ``external'' classes is stored as \texttt{mcccadata\$C} and can also be found in the output of \texttt{summary()}. For this example, the number of classes is $C=4$, and we therefore need to define an integer vector of length 4 containing the number of class-specific clusters that we aim to retrieve. 

The numbers of the class-specific clusters need to be supplied by the user. As in most cluster methods, this choice is not-trivial. The mccca package contains several functions that can be used to make a choice. However, as the mealDrink data were generated according to a specific cluster structure, we can use the ``true'' (i.e., the numbers used to generate the data) numbers for the clusters here. These true numbers are: $2$ (American females), $1$ (American males), $2$ (Japanese females) and $2$ (Japanese males). After specifying this as \texttt{K.vec}, we perform MCCCA:
%
\begin{CodeChunk}
\begin{CodeInput}
R> K.vec=c(2,1,2,2)
R> set.seed(5)
R> res=MCCCA(mcccadata,K.vec=K.vec)
\end{CodeInput}
\end{CodeChunk}

The \texttt{MCCCA()} function requires an MCCCAdata object (i.e., a list produced by\\
\texttt{create.MCCCAdata()}) and a vector with class-specific numbers of clusters  \texttt{K.vec}. Table~\ref{tab:MCCCAfunc} provides a complete overview of the available input arguments and output fields.

We can use \texttt{plot()} to visualize the results of the MCCCA analysis. The resulting biplot includes labels for the class-specific clusters. By default, these labels are formed by combining the categories of the external variables (e.g., ``American\&female''). However, the labels may be customized using the \texttt{classlabel} field. For our example, we create short labels by only using the first letter of each class:

\begin{CodeChunk}
\begin{CodeInput}
R> shortlabels=c("AF","AM","JF","JM")
R> plot(res,classlabel=shortlabels)
\end{CodeInput}
\end{CodeChunk}

\begin{figure}[t!]
\centering
\includegraphics[width=15cm]{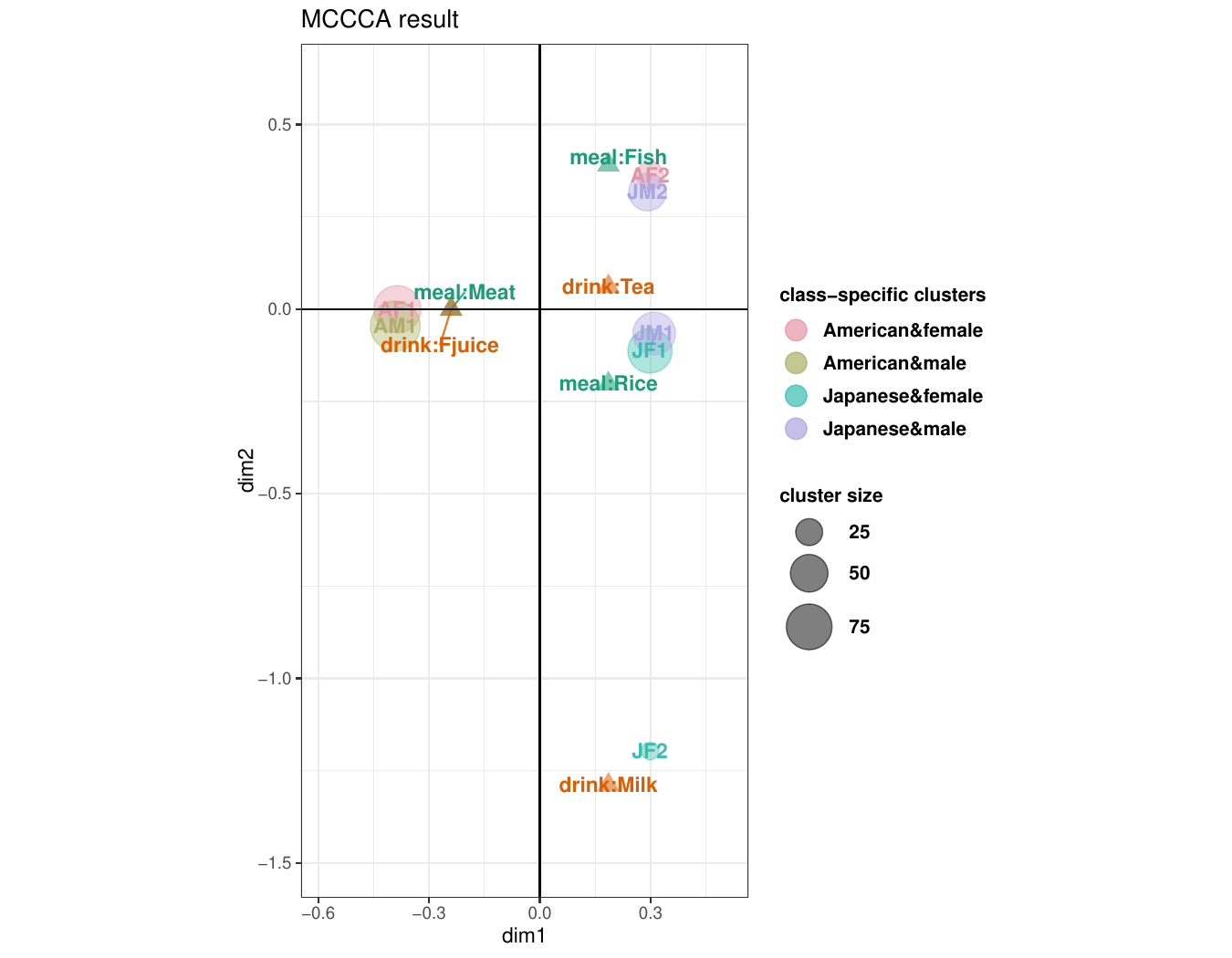}
\caption{\label{fig:mccca_mealplot1}Results using MCCCA for the \texttt{mealDrink} example. Class-specific cluster points (i.e., bubbles) are labeled as ``AM'' (American males), ``JM'' (Japanese males), ``AF'' (American females), and ``JF'' (Japanese females) followed by a cluster number. The smaller the number, the larger the number of observations belonging to the cluster. The bubble sizes correspond to cluster sizes. The triangle points correspond to category points and are labeled using the appropriate combinations of variable names and category labels.}
\end{figure}

The resulting plot is presented in Figure \ref{fig:mccca_mealplot1}. Different colors and symbols are used to distinguish between categories of the active variables and class-specific clusters. In particular, triangle points correspond to categories of the active variables, and categories belonging to different variables are displayed using different colors. For the class-specific cluster points, colors indicate the classes. Moreover, each class-specific cluster center is represented by a bubble. The bubble sizes are proportional to the cluster sizes, and cluster numbers are assigned according to relative cluster size, where a value of $1$ indicates the largest cluster within a class. A legend showing the class-color correspondence and a scale for the bubble sizes is added automatically.

Note that the order of \texttt{shortlabels} should be the same as the order of \texttt{classlabel} in the \texttt{mcccadata} object (i.e., \texttt{mcccadata\$classlabel}). There are many other options for customization of the visualization using the \texttt{plot()} function. See Table $\ref{tab:plotfunc}$ for a complete overview.

\begin{table}[t!]
	\centering
	\caption{{\small List of \texttt{MCCCA()} arguments and outputs.}}
	\scalebox{0.75}{ 
			\begin{tabular}{ll}
				\hline
				    Arguments & Description\\
                    \hline
				\texttt{mcccadata} & A list created by \texttt{create.MCCCAdata()}.\\
				\texttt{K.vec} & An integer vector of length $C$. Each element corresponds to \\
    & the number of clusters in each class, which needs to be specified for estimation.\\
    \texttt{nstart} & An integer indicating the number of random initial values. The default is 30.\\
		\texttt{ndim} & An integer indicating the dimension of quantification (i.e., $p$). The default is 2.\\
		\texttt{maxit} & An integer indicating the maximum number of iterations.\\
		\texttt{tol} & A numeric value indicating the absolute convergence tolerance. The default is 1e-6. \\
            \texttt{update.kmeans} & A logical value, indicating whether a cluster indicator matrix is updated by $k$-means or\\
           & referring to the distance. The default is \texttt{FALSE} (not using $k$-means).\\
            \texttt{verbose} & A logical value. If \texttt{TRUE}, tracing information on the progress of the optimization is produced.\\
				\hline
    Output & Description\\
    \hline
    \texttt{G} & A $K\times p$ quantification matrix for the clusters.\\
    \texttt{B} & A $Q\times p$ quantification matrix for the categories.\\
    \texttt{Fn} & An $N\times p$ quantification matrix for the observations.\\
    \texttt{Gg}, \texttt{Bg}, \texttt{Fng} & 
    Re-scaled versions of \texttt{B},\texttt{G} and \texttt{Fn} respectively, where the average squared deviations\\
    & from the origin for the row and column points is the same\\
    & (See section 2.3 in \cite{1} for  details).\\
    & These coordinates are used in the \texttt{plot} by default.\\
    \texttt{cluster.class.index} & An integer vector of length $K$, showing the class index to which each cluster corresponds.\\
    \texttt{cluster.index}& An integer (from $1:K$) vector of length $K$, showing the cluster index \\
    & to which each cluster corresponds.\\
    \texttt{cluster.label.n.vec} & A character vector of length $N$, where each element represents \\
    & the cluster label for each observation.\\
    \texttt{cluster.label.n.list}& A list of $C$ character vectors, giving the cluster labels for each observation in each class.\\
    \texttt{cluster.index.n.vec}& An integer vector of length $N$, showing the same information \\
    & as \texttt{cluster.label.n.vec} but organized by cluster index.\\
    \texttt{cluster.index.n.list}& A list of $C$ vectors, showing the same information as \texttt{cluster.label.n.list}\\
    & but organized by cluster index.\\
    \texttt{optval} & A numeric value indicating the optimal value of the objective function.\\
    \texttt{catename.vec} & A character vector of length $Q$ with the category names as labels\\
    & for all categories of the active variables.\\
    \texttt{catename.vari.vec} & A character vector of length $Q$ with combinations of active variable names \\
    & and categories \texttt{catename.vec} (by default, this is used as the column name of \texttt{B} and \texttt{Bg}).\\
    \texttt{cate.removed} & If there is a category that is not chosen by any observation, this gives which category\\
    & was removed (given by the index of column in dummy matrix). Otherwise, return \texttt{NULL}.\\
    \texttt{inertia} & A table containing the adjusted inertias and their relative and cumulative values.\\
    \texttt{inertia.G} & A table containing the adjusted inertias and their relative and cumulative values \\
    & based on Greenacre's normalization.\\
    \texttt{q.vec} & A copy of \texttt{q.vec} in the \texttt{mcccadata} object.\\
    \texttt{K.vec} & A copy of \texttt{K.vec} in the argument of \texttt{MCCCA()}.\\
    \texttt{classlabel} & A copy of \texttt{classlabel} in the \texttt{mcccadata} object.\\
                    \hline
            
			\end{tabular}
	}
	\label{tab:MCCCAfunc}
\end{table}

\begin{table}[t]
    \centering
	\caption{{\small List of \texttt{plot()} arguments.}}
	\scalebox{0.75}{ 
			\begin{tabular}{ll}
				\hline
				    Arguments & Description\\
                    \hline
				\texttt{res} & A \texttt{mccca} object, a list created by \texttt{MCCCA}.\\
				\texttt{main} & A character string giving the title of biplot.  The default is "MCCCA result".\\
		\texttt{xlim} & A numeric vector of length 2 giving the plotting range for the $x$ (horizontal) axis. \\
  & If \texttt{NULL}, the range is automatically determined.\\
		\texttt{ylim} & A numeric vector of length 2 giving the plotting range for the $y$ (vertical) axis\\
            & (the same role as \texttt{xlim}).\\
		\texttt{plot.ind} & If \texttt{TRUE}, coordinates of observations are also plotted. The default is \texttt{FALSE}.\\
            \texttt{classlabel} & A character vector of length $C$ ($C$:the number of classes) giving labels for each class \\
            & to be displayed on the biplot. If \texttt{NULL}, \texttt{classlabel} saved in \texttt{mcccadata} is used.\\
            \texttt{classlabel.legend} & A character vector of length $C$ giving labels for each classes to be shown on the legend. \\
            & If \texttt{NULL}, \texttt{classlabel} is used for the legend.\\
            \texttt{catelabel}& A character vector of length $Q$ ($Q$: the total number of categories) giving labels \\
            & for each category to be displayed on the biplot. If \texttt{NULL}, \texttt{rownames(B)} is used.\\
            \texttt{variname} & A character vector of length $J$ ($J$:the number of active variables) giving labels \\
            & for each variable to be displayed on the biplot. If \texttt{NULL}, the variable names \\
            & included in the row names of \texttt{B} or \texttt{Bg} are used.\\
            \texttt{include.variname} & If \texttt{TRUE}, the variable name is included in the category label on the biplot\\
            & (e.g., the category point named "male" for the variable named "v1" \\
            & is displayed as "v1:male" on the biplot). The default is \texttt{TRUE}.\\
            \texttt{show.onlyminmax} & If \texttt{TRUE}, only the bubbles corresponding to the maximum and minimum cluster sizes are\\
            & displayed on the legend. The default is \texttt{FALSE}.\\
            \texttt{break.size} & A numeric vector that adjusts the size of bubble displayed on the legend.\\
            \texttt{scale.gamma} & If \texttt{TRUE}, quantifications are scaled such that the average squared deviation from the origin \\
            & of the row and column points is the same. The default is \texttt{TRUE}.\\
            \texttt{scatter.level} & A numeric value that adjusts the scatter of points on the biplot. \\
            & Larger values spread the labels farther apart to reduce overlap. The default is 2.\\
            \texttt{txtsize} & A numeric value that adjusts the text size of labels on the biplot. The default is 3.8.\\
            \texttt{alp.trans} & A numeric value from 0 to 1 which adjusts the transparency of the bubble point. \\
            & The default is 0.4.\\
            \texttt{show.inertia} & If \texttt{TRUE}, the axis labels show the percentages of adjusted inertia\\
            & obtained by applying MCA to the original data. The default is \texttt{FALSE}.\\
            \hline
			\end{tabular}
	}
	\label{tab:plotfunc}
\end{table}

To inspect numerical results, in particular coordinates and fit per dimension, one can use \texttt{summary()}. 
\begin{CodeChunk}
\begin{CodeInput}
R> summary(res)
The number of clusters specified for each class:
American&female   American&male Japanese&female   Japanese&male 
              2               1               2               2 

Coordinates (gamma scaled) and sizes of each class-specific cluster:
                   dim1   dim2 sizes relative sizes
American&female1 -0.386  0.000    80          0.762
American&female2  0.298  0.363    25          0.238
American&male1   -0.386  0.000    90          1.000
Japanese&female1  0.298 -0.113    70          0.875
Japanese&female2  0.298 -1.197    10          0.125
Japanese&male1    0.298 -0.113    65          0.565
Japanese&male2    0.298  0.363    50          0.435

Coordinates of each category (gamma scaled):
               dim1   dim2
meal:Fish     0.186  0.391
meal:Meat    -0.240  0.000
meal:Rice     0.186 -0.202
drink:Fjuice -0.240  0.000
drink:Milk    0.186 -1.289
drink:Tea     0.186  0.061
\end{CodeInput}
\end{CodeChunk}

The output of \texttt{summary} consists of the estimated coordinates of class-specific cluster centroids and active variable category points, i.e., \texttt{Gg} and \texttt{Bg}, respectively (see also Table \ref{tab:MCCCAfunc}). Note that the ``gamma-scaled'' coordinates are scaled in such a way that the average squared deviation from the origin of the row and column points is the same (See section 2.3 in \cite{1} for details). Furthermore, ``American\&female1'' indicates the largest cluster in the ``American and female'' class. In the table with cluster centroid coordinates, cluster sizes and relative cluster sizes within each class are also listed. Here, we see, for example, that 10 observations belong to the smaller cluster of the ``Japanese female class''.

To assess the quality of the two dimensional visualization,  the relative percentages of adjusted inertias, as proposed by \cite{5} (also described in \cite{6}), are provided along the appropriate dimensions in Fig. \ref{fig:mccca_mealplot1}. More detailed information on the distribution of inertia is stored in the \texttt{inertia} field. These adjusted inertias are obtained by applying MCA to the full dataset. That is, for our example:

\begin{CodeChunk}
\begin{CodeInput}
R> res$inertia
      adInertias relative_adInertias cum_adInertias
dim 1 1.00000000          0.97596154      0.9759615
dim 2 0.02463054          0.02403846      1.0000000
\end{CodeInput}
\end{CodeChunk}

Here, the first column gives the adjusted, the second the relative, and the third the cumulative inertia. For these example data, the cumulative adjusted inertia reaches 1.0 at the second dimension, because the remaining dimensions have adjusted inertia equal to zero after Benzécri's adjustment.

\subsubsection*{The averaging approach}

To further illustrate the use of MCCCA, we can apply the averaging approach mentioned in Section $\ref{sec:intro}$ and compare results. As the averaging approach corresponds to MCCCA with the numbers of all class-specific clusters set to $1$, we can achieve this by executing:
\begin{CodeChunk}
\begin{CodeInput}
R> K.vec=c(1,1,1,1)
R> res.ave=MCCCA(mcccadata,K.vec=K.vec)
\end{CodeInput}
\end{CodeChunk}

We can plot the results again by using \texttt{plot(res.ave)}.
\begin{CodeChunk}
\begin{CodeInput}
R> plot(res.ave,classlabel=shortlabels,main="Averaging result")
\end{CodeInput}
\end{CodeChunk}

\begin{figure}[t!]
\centering
\includegraphics[width=10cm]{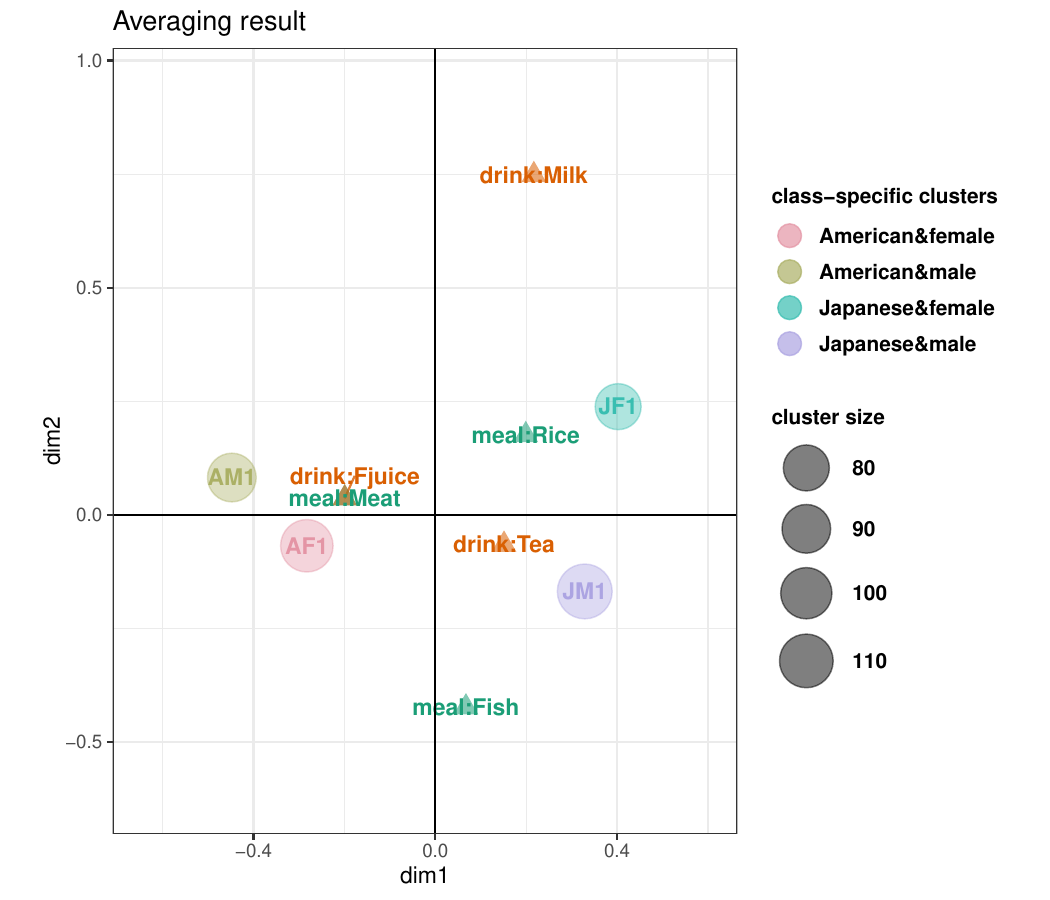}
\caption{\label{fig:mccca_mealplot2} Results using the averaging approach for \texttt{mealDrink} data. No class-specific clusters are obtained and all points have label 1 and the bubble sizes correspond to the overall class sizes.}
\end{figure}

The resulting plot can be seen in Figure \ref{fig:mccca_mealplot2}. Comparing the averaging approach (Fig. \ref{fig:mccca_mealplot2}) with the MCCCA results (Fig. \ref{fig:mccca_mealplot1}) shows that the averaging approach only reveals a single association pattern for each class, whereas MCCCA can identify heterogeneous tendencies \emph{within} classes. For example, the averaging approach associates the Japanese female class with ``Rice'' and ``Tea''. In contrast, MCCCA reveals two distinct clusters within this class: a larger cluster associated with ``Rice'' and Tea'' and a smaller cluster associated with ``Rice'' and ``Milk''. Similarly, for American females, the two clusters differ with respect to meal preferences (i.e., Fish or Meat).

\subsection{Roadside Accidents and Conditions}

A data set obtained from the U.K. Department for Transport's road safety statistics (\url{https://www.gov.uk/government/collections/road-safety-statistics}) concerning road accidents is available in the \textbf{mccca} package. In particular, the \texttt{accident} data set consists of a selection of accidents concerning one casualty (either a driver or pedestrian) that occurred in January 2016, with at most two parties involved. The resulting data contains $N=3026$ accidents (observations). For each accident, there are $6$ categorical variables. We use the first four: ``lighting conditions'', ``weather conditions'', ``road surface conditions'', and ``speed limit'', as active variables. The remaining two: ``casualty class'' and ``area'' are treated as external variables. A complete list of these variables and their categories is given in Table $\ref{tab:catename}$.  

As both external variables consist of two categories, a total of four classes can be distinguished: ``driver and urban'', ``driver and rural'', ``pedestrian and urban'' and ``pedestrian and rural''. The aim for these data is to determine how conditions corresponding to car accidents vary depending on the type of accident as identified by the four ``classes'' defined above. As before, we load the data and specify the active and external variables: 
\begin{CodeChunk}
\begin{CodeInput}
R> data(accident)
R> active<-c(1:4)
R> external<-c(5:6)
R> dat.act<-accident[,active]
R> dat.ext<-accident[,external]
\end{CodeInput}
\end{CodeChunk}
\begin{table}
	\centering
	\caption{{\small Categories for each variable and their corresponding labels in biplots and descriptions.}}
	\scalebox{0.8}{
		\begin{tabular}{llll}
			\hline
			Variable type & Variable name & Label & Description\\
			\hline
			Active variables & Light conditions & L0 & Daylight\\
			&  & L1 & Darkness: street lights present and lit \\
			&  & L2 & Darkness: street lights present but unlit \\
			&  & L3 & Darkness: no street lighting \\
			& Weather conditions & Fine & Fine without high winds \\
			&  & Rain & Raining without high winds \\
			&  & Snow & Snowing without high winds \\
			&  & Fine\_w & Fine with high winds \\
			&  & Rain\_w & Raining with high winds \\
			&  & Snow\_w & Snowing with high winds \\
			&  & Fog & Fog or mist — if hazard \\
			&  & Other & Other \\
			& Road surface conditions & Dry & Dry \\
			&  & Wet & Wet / Damp \\
			&  & Snow & Snow \\
			&  & Frost & Frost / Ice \\
			&  & Flood & Flood (surface water over 3cm deep) \\
			& Speed limit & S30 & Speed limit is up to 30km/h\\
			&  & S70 & Speed limit is from 30km/h up to 70km/h\\
			External variables & Casualty class & Driver & Casualty is one driver\\
&  & Ped & Casualty is one pedestrian\\
& Area & Urban & Occurring in urban area\\
&  & Rural & Occurring in rural area\\
			\hline
		\end{tabular}
	}
	\label{tab:catename}
\end{table}

Similar to the meal-drink example, we create an \texttt{mcccadata} object using \texttt{create.MCCCAdata()}. This shows all existing classes and their sizes:

\begin{CodeChunk}
\begin{CodeInput}
R> mcccadata=create.MCCCAdata(dat.act,dat.ext)
1th class: (Driver,Rural) data, n=818
2th class: (Driver,Urban) data, n=305
3th class: (Ped,Rural) data, n=230
4th class: (Ped,Urban) data, n=1673
The total number of classes: 4 
\end{CodeInput}
\end{CodeChunk}
To apply MCCCA, the numbers of class-specific clusters need to be specified. In our previous example, in Section $\ref{mealDrink}$, we used the ``true'' number of clusters. Here, and in most real data applications, the number of clusters is unknown. Determining an appropriate number is, as in any clustering method, a difficult task as there is usually no objective validation measure available. In fact, MCCCA is an unsupervised joint clustering and visualization method. Hence, interpretability of results is an important, but difficult to objectively quantify, criterion. 

In cluster analysis there are many ways for selecting the number of clusters and all of these can also be applied for MCCCA. It is beyond the scope of this paper to review such methods. However, the \textbf{mccca} package provides the function \texttt{decideK()} to assist in determining the number of clusters. In particular, \texttt{decideK()} performs $k$-means clustering to MCA results (i.e., the row coordinates) based on observations for each class, using the active variables, for a specified range of values for the number of clusters. It calculates the value for a specified cluster quality index for each number of clusters in the range, and uses these to suggest a choice for the number of clusters. The function requires an \texttt{mcccadata} object, a range of values, indicated by its minimum and maximum, for the number of clusters, and a cluster quality index. Here, similar to \cite{1}, the default cluster quality measure is the Krzanowski–Lai index ($KL$: \cite{7}).  However, three other cluster quality indicators: $CH$: Calinski \& Harabasz index \cite{8}; $S$: Silhouette index \cite{9} and $DB$: Davies-Bouldin's index \cite{10} are also available. A different index can be selected using the \texttt{cindex} argument of \texttt{decideK()}.

To illustrate, we consider a range for the number of clusters from $2$ to $8$:
\begin{CodeChunk}
\begin{CodeInput}
R> Krange=c(2,8)
R> set.seed(5)
R> KLres=decideK(mcccadata,Krange=Krange)
\end{CodeInput}
\end{CodeChunk}
Here the number of clusters to be considered is between $2$ and $8$, and \texttt{decideK} considers the $KL$ index for each choice of $K$ in each class and stores the results. To inspect the results, we use \texttt{summary}:
\begin{CodeChunk}
\begin{CodeInput}
R> summary(KLres)
The Krzanowski & Lai index values for each cluster in each class:
                   clusters=2 clusters=3 clusters=4 clusters=5 clusters=6
Driver&Rural class      0.188       4.94      0.848      1.748      1.837
                   clusters=7 clusters=8
Driver&Rural class      0.616      3.628
                   clusters=2 clusters=3 clusters=4 clusters=5 clusters=6
Driver&Urban class      0.034      4.524      2.133      0.937      1.523
                   clusters=7 clusters=8
Driver&Urban class      1.184      0.831
                clusters=2 clusters=3 clusters=4 clusters=5 clusters=6
Ped&Rural class      1.035      1.493     26.213      0.087      0.804
                clusters=7 clusters=8
Ped&Rural class       0.81      1.255
                clusters=2 clusters=3 clusters=4 clusters=5 clusters=6
Ped&Urban class      0.312      4.553      0.552      3.154      1.934
                clusters=7 clusters=8
Ped&Urban class      0.736      0.719

The best number of clusters for each class:
Driver&Rural class Driver&Urban class    Ped&Rural class    Ped&Urban class 
                 3                  3                  4                  3 
\end{CodeInput}
\end{CodeChunk}
This output shows the best, according to the KL index, number of clusters for each class\footnote{Note that these results are not the same as those published in \cite{1} due to a difference in the number of random starts. In particular, the \texttt{decideK()} functions uses 1000 random starts, whereas for the application in the paper, 100 random starts were used. Note, however, that the overall interpretation of the results does not change.}.

In addition to the numerical results obtained using \texttt{summary()}, we can obtain a more detailed picture of internal cluster qualities for different numbers of clusters by visualizing the \texttt{decideK()} results using \texttt{plot()}:

\begin{CodeChunk}
\begin{CodeInput}
R> plot(KLres)
\end{CodeInput}
\end{CodeChunk}

\begin{figure}[t!]
\centering
\includegraphics[width=14cm]{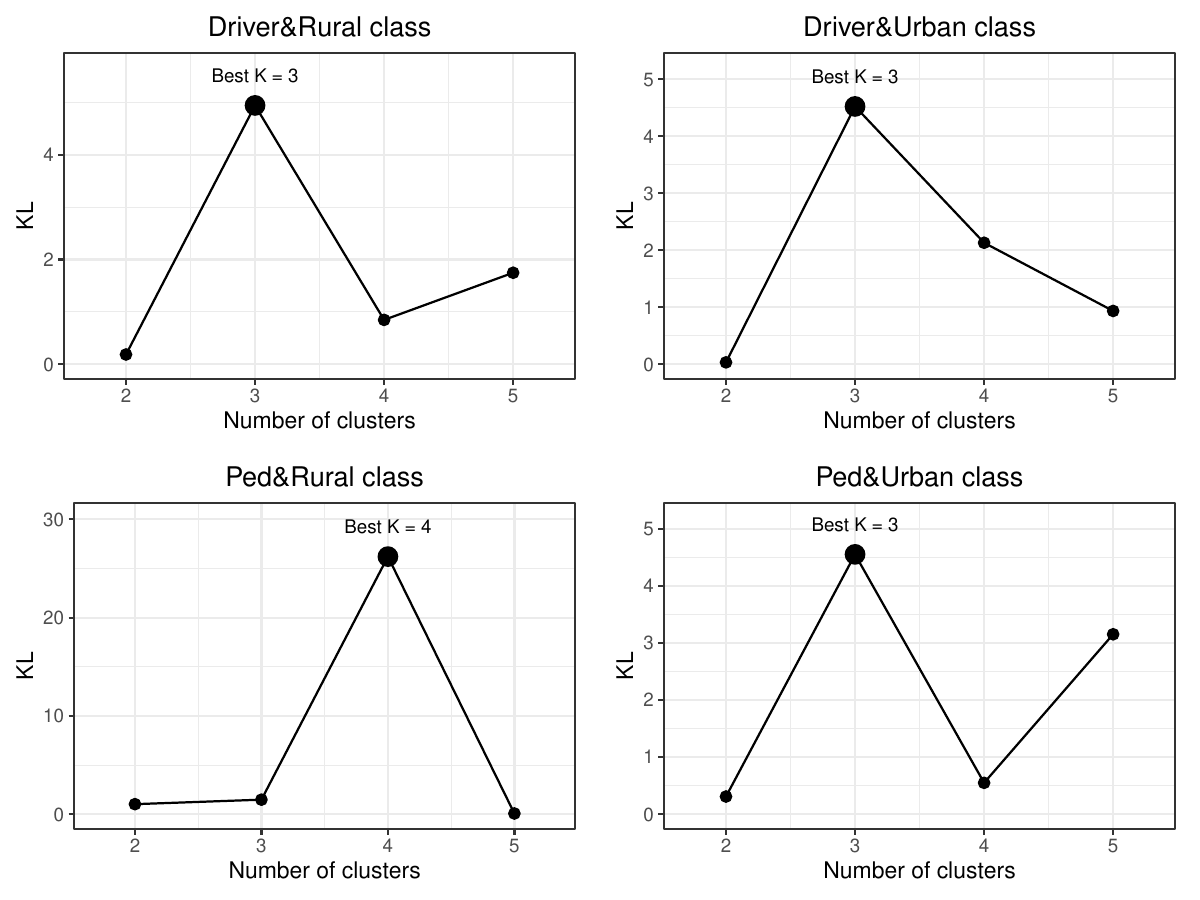}
\caption{\label{fig:decideKplot} Visualization of the results obtained by \texttt{plot(KLres)}. Each panel corresponds to one class. The horizontal axis represents the candidate number of clusters, and the vertical axis represents the corresponding values of the KL index. The larger point indicates the selected number of clusters for each class.}
\end{figure}

Figure~\ref{fig:decideKplot} shows all four plots together. 

The suggested numbers of class-specific clusters are stored as \texttt{cls.best} in the \texttt{decideK} object. We can directly use this as an input argument in the \texttt{MCCCA} function:
\begin{CodeChunk}
\begin{CodeInput}
R> K.vec=KLres$cls.best
R> K.vec
   Driver&Rural class Driver&Urban class    Ped&Rural class    Ped&Urban class 
                 3                  3                  4                  3
R> set.seed(5) #to produce the same result as in this paper.
R> res=MCCCA(mcccadata,K.vec=K.vec)
\end{CodeInput}
\end{CodeChunk}


Directly plotting the results using \texttt{plot}, yields a plot that is hard to interpret due to the lengthy labels for the active variables. However, we can customize labels for variables using the \texttt{variname} argument of the \texttt{plot} function. For example, we can replace the variable names by their first letters:
\begin{CodeChunk}
\begin{CodeInput}
R> plot(res,variname=c("L","W","R","S"))
\end{CodeInput}
\end{CodeChunk}
\begin{figure}[t!]
\centering
\includegraphics[width=15cm]{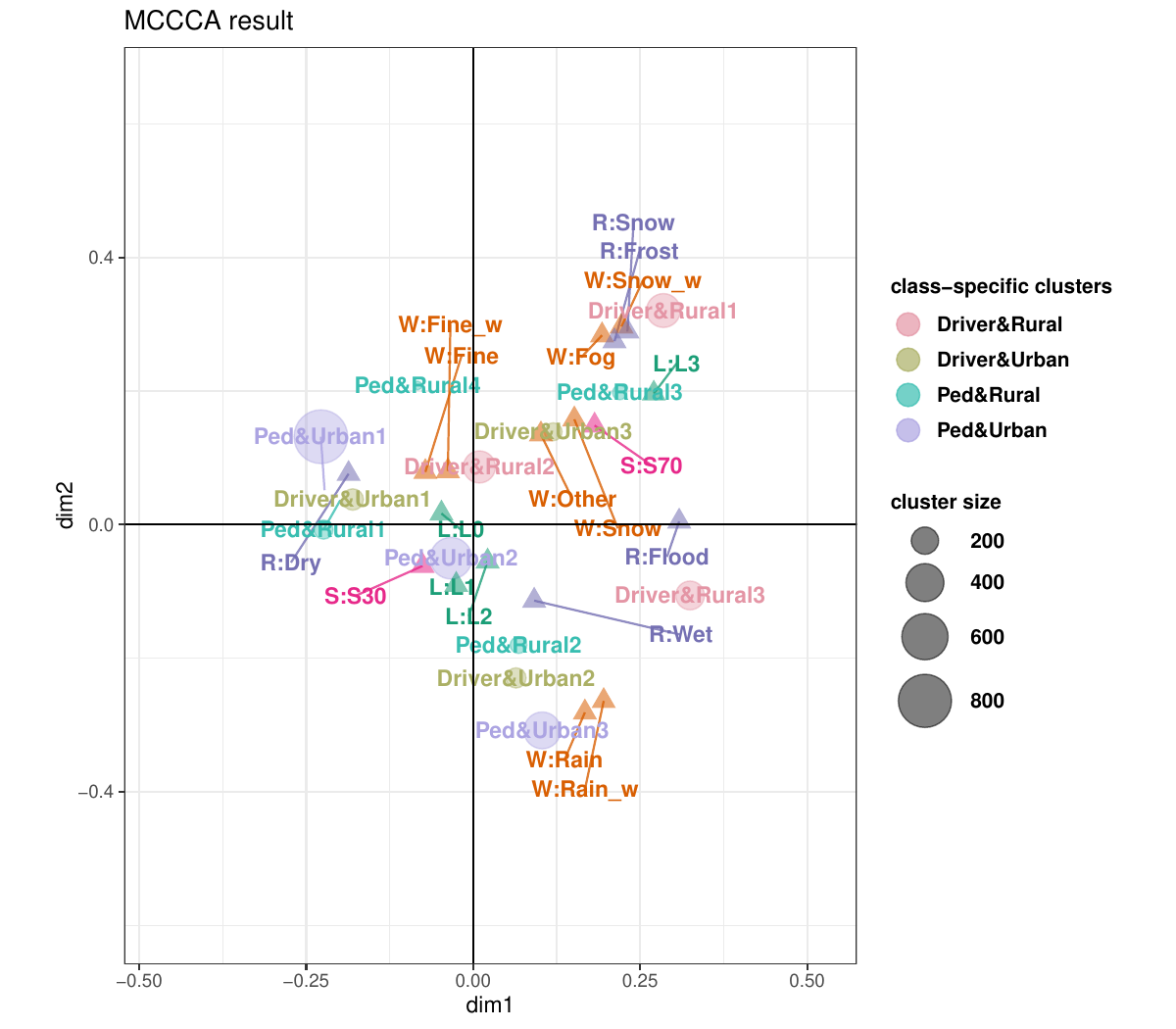}
\caption{\label{fig:accires} Results using MCCCA for \texttt{accident} data. The numbered bubbles indicate cluster points; the smaller the number, the greater the number of observations belonging to that cluster. Other character labels correspond to categories, and at the beginning of each category is the abbreviated variable name for that category (e.g., ``R:Flood'' indicates the ``Flood'' category for the ``road surface conditions'' variable).}
\end{figure}

The configuration of the category quantifications shows that there are roughly three types of accidents. On the positive side of the horizontal axis, there are ``Snow (weather and road surface conditions)'', ``Fog (weather conditions)'' and ``L3 (darknesss; light conditions)''. All of these concern ``bad'' driving conditions. On the negative side we find ``Fine (weather conditions)'', ``Dry (road surface conditions), and ``L0 (daylight; light conditions)'', all of which indicate ``good'' driving conditions. Furthermore, the second dimension appears to separate ``rainy'' conditions from the rest. In particular, on the negative side we find categories as ``Rain (weather conditions)'' and ``Wet (road surface conditions)''.  

Around categories corresponding to ``good'' driving conditions, we find the largest ``pedestrian and urban'', ``pedestrian and rural'' and ``driver and urban'' class-specific clusters. Hence, accidents in these classes tend to occur relatively often (i.e., more than on average) in ``good'' conditions. This may partly reflect the fact that favorable driving conditions occur more frequently than adverse conditions and therefore account for a larger number of accidents overall. Nevertheless, this indicates that accidents that occur in spite of favorable conditions, often involve pedestrians, and, in urban areas, they often involve drivers. 

On the other hand, on the right side of the plot we find a large cluster in the ``Driver and Rural'' class indicating a strong association with ``bad'' driving conditions. In rural areas (where there are relatively few pedestrians), accidents involving drivers are more likely to occur under ``bad'' conditions. Finally, around the categories corresponding to ``rainy'' conditions, we find smaller clusters for all classes. This indicates that some accidents occurred in rainy conditions for all classes. 

By looking at the class-specific clusters we can distinguish whether there are groups of observations that differ with respect to their association with the active variables. The relative explained inertias, plotted along the dimensions in Fig. \ref{fig:accires}, indicate that the two dimensional visualization accounts for 80.2\% of the inertia.

\section{Conclusion} \label{sec:conclu}

In this paper, we presented the \textsf{R} package \textbf{mccca}, that implements MCCCA, a non-trivial joint dimension reduction and cluster analysis method proposed in \cite{1}. As MCCCA is primarily a descriptive data analysis method, an easy-to-use implementation is essential. Such an implementation did not yet exist, limiting the potential use of the method. The \textbf{mccca} package implements all MCCCA estimation procedures, as well as tuning functions for selecting appropriate numbers for the clusters, and customizable visualization functions that allow for the depiction of the various aspect of MCCCA solutions.

To apply MCCCA, users first need to indicate the two types of variables: active and external. Based on the external variables, observations are separated into multiple ``classes'', after which class-specific clusters, based on the active variables, are obtained. We presented a typical ``flow of analysis'' using two examples illustrating the workings of both the method and the options available in the package. MCCCA is an exploratory data analysis method, and the new \textbf{mccca} package enables users to estimate and assess, different models (using different parameter settings) both numerically and visually. 

Finally, MCCCA is a method for categorical data. It would be interesting to consider implementations that involve active numerical variables. However, this requires methodological innovations and non-trivial visualization choices. We leave this for future work.





\section*{Acknowledgments}

This work was supported by JSPS KAKENHI Grant Number JP20K19755.

\begin{leftbar}
\end{leftbar}

\end{document}